\documentclass[superscriptaddress,10pt,prx,twocolumn,showpacs,longbibliography,float fix]{revtex4-2}
\usepackage{blindtext}
\usepackage{lipsum}
\usepackage{graphics}
\usepackage{amsmath}
\usepackage{graphicx}
\usepackage{colortbl}
\usepackage{amsthm}
\usepackage{graphics}
\usepackage{amssymb}
\usepackage{amsthm}
\usepackage{verbatim}
\usepackage{physics}
\usepackage{float}
\usepackage{mathtools}
\usepackage{microtype}
\usepackage[normalem]{ulem}
\usepackage[colorlinks=true,citecolor=blue,linkcolor=red,urlcolor=blue,pdftitle={Hierarchy cost}]{hyperref}
\usepackage[dvipsnames]{xcolor}
\definecolor{darkblue}{rgb}{0.,0.,0.4}

\def\bea{\begin{eqnarray}}
\def\eea{\end{eqnarray}}

\usepackage{array}
\usepackage{xfrac}

\usepackage{mathrsfs}

\usepackage{soul}
\usepackage{tablefootnote}
\usepackage{mathtools}
\usepackage{tikz}

\newtheorem{theorem}{Theorem}

\newtheorem{proposition}{Proposition} 

\def\dimer{\tikz[baseline=-0.5ex]{
			\fill (0,0) circle (1.5pt) coordinate (A);
			\fill (3.0ex, 0) circle (1.5pt) coordinate (B);
			\draw[black, thick] (A)--(B);}
}

\def\twodimer{\tikz[baseline=-0.5ex]{
		\fill (0,0) circle (1.5pt) coordinate (A);
		\fill (3.0ex, 0) circle (1.5pt) coordinate (B);
		\fill (6.0ex,0) circle (1.5pt) coordinate (C);
		\fill (9.0ex, 0) circle (1.5pt) coordinate (D);
		\draw[black, thick] (B)--(C);
		\draw[black, thick] (A) .. controls (3.0ex, 1.5ex) and (6.0ex, 1.5ex) .. (D);
	}
    }

\begin{document}

\title{Entanglement cost hierarchies in quantum fragmented mixed states}

\author{Subhayan Sahu}
\affiliation{ Perimeter Institute for Theoretical Physics, Waterloo, Ontario N2L 2Y5, Canada}%

\author{Yahui Li}
\affiliation{
Technical University of Munich, 
TUM School of Natural Sciences, 
Physics Department,
James-Franck-Str. 1,
85748 Garching,
Germany}%
\affiliation{Munich Center for Quantum Science and Technology (MCQST), Schellingstr. 4, 80799 M{\"u}nchen, Germany}

\author{Pablo Sala}%
\affiliation{Department of Physics and Institute for Quantum Information and Matter, California Institute of Technology, Pasadena, CA 91125, USA}
\affiliation{Walter Burke Institute for Theoretical Physics, California Institute of Technology, Pasadena, CA 91125, USA}

\begin{abstract} 

Strong symmetries enforce non-trivial quantum entanglement patterns on the stationary states of symmetric open quantum dynamics. 
Specifically, non-commuting conserved quantities lead to long-range quantum entanglement even for infinite temperature mixed states within fixed symmetry sectors. 
Leveraging the commutant algebra framework, we show that various bipartite entanglement measures for mixed states---including exact and asymptotically-exact entanglement costs and squashed entanglement, which are generally intractable for a generic many-body mixed state---can be computed for this class of states.
In particular, we focus on strongly symmetric maximally mixed states arising from the Temperley-Lieb model, which features quantum Hilbert space fragmentation with exponentially large (in system size) non-Abelian commutants. 
We find that while both the logarithmic negativity and the `exact' entanglement cost for equal-size bipartitions scale with the volume of the system, the entanglement of formation, squashed entanglement, entanglement cost, and distillable entanglement exhibit subextensive scaling. 
We relate this separation in entanglement measures to a parametric difference between the entanglement cost of exact and asymptotically-exact state preparations, and infer this to be a consequence of a particular pattern of quantum Hilbert space fragmentation. 

\end{abstract}

\maketitle

\textbf{Introduction---} While the von Neumann entropy is the unique good entanglement measure for pure states, characterizing entanglement in mixed quantum many-body systems turns out to be much harder~\cite{horodecki_quantum_2009}. Yet, mixed states naturally arise in realistic settings, e.g., at any non-zero temperature~\cite{Hastings_2011}, in a noisy quantum device~\cite{Preskill_2018}, or in dissipative and monitored dynamical setups~\cite{Fisher_2023, Potter_2022}. Therefore, characterizing entanglement in mixed quantum many-body states is an important question which remains rather open.

\renewcommand{\arraystretch}{1.4}
\setlength{\arrayrulewidth}{0.3mm}
\begin{figure}[t!]
    \centering
\begin{tikzpicture}
\node at (-0.5,-2.2){\includegraphics[scale=0.8]{{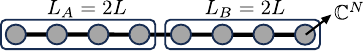}}};
\node at (-4.5,1.3) {(a)};
\node at (-4.5,-2.2) {(b)};

\node at  (-0.5,0.){    

    \begin{tabular}{c|cccc|cc}
        & \multicolumn{4}{c|}{$\mathscr{E}_{(A:B)}^{<}$} & \multicolumn{2}{c}{$\mathscr{E}_{(A:B)}^{>}$}\\
        \hline\hline
        $\mathcal{C}(L)$& \multicolumn{4}{c|}{$\sum\limits_{\lambda} p_\lambda \log(d_\lambda)$} & \multicolumn{2}{c}{$\log\left(\sum \limits_{\lambda } p_\lambda d_\lambda\right)$} \\ \hline 
        \multicolumn{7}{c}{$\rho_{\text{MMIS}}$ in Eq.~\eqref{eq:MMIS_def} for $L_A =L_B = 2L$} \\ \hline
        $\mathcal{C}_{\textrm{su}(N)}(L)$ & \multicolumn{4}{c|}{$\sim  c_N\log(L)$} & \multicolumn{2}{c}{$\sim c_N\log(L)$}   \\ \hline
        $\mathcal{C}_{\textrm{RS}(N\geq 3)}(L)$ & \multicolumn{4}{c|}{$\sim \log(q_N) \sqrt{2L}$} & \multicolumn{2}{c}{$\sim \frac{\log^2(q_N)}{2}L$}  
    \end{tabular}
};
    
\end{tikzpicture}
    \caption{\textbf{Bipartite entanglement measures for maximally mixed invariant states.} (a) Exact expressions for various entanglement measures for the states defined in Eq.~\eqref{eq:singlet_mixed_def} for a commutant algebra $\mathcal{C}(L)$. We prove that $ E^{D}_{(A:B)} =  E^{\mathrm{sq}}_{(A:B)} =  E^{C}_{(A:B)} =  E^{F}_{(A:B)} \equiv \mathscr{E}_{(A:B)}^{<} $, and $E^{N}_{(A:B)} = E^{C,\mathrm{PPT,exact}}_{(A:B)} \equiv \mathscr{E}_{(A:B)}^{>}$; $d_\lambda$ is the dimension of the irreducible representation $\lambda$ of $\mathcal{C}(L)$. The last two rows show $\mathscr{E}_{(2L:2L)}^{>}$ and $\mathscr{E}_{(2L:2L)}^{<}$ for the maximally mixed invariant state $\rho_{\text{MMIS}}$ in Eq.~\eqref{eq:MMIS_def} defined on a $4L$- sized chain of $N$-dits, with the spatial half-partition depicted in (b), and for the commutants corresponding to SU$(N)$ and Read-Saleur (RS) algebras (for Temperley-Lieb model). For SU$(N)$, both scale as $\log L$ with same $N-$dependent prefactor $c_N$, but with different $O(1)$ additive corrections, while for RS, they have parametric separation. The $q_N$ in the last row is given by $N=q_N+q_N^{-1}$ with $q_N\geq 1$.
    }
    \label{tab:summary_measures}
\end{figure}

The amount of entanglement in a given mixed state depends crucially on the details of the operational protocol, but most of the desirable measures of entanglement are practically impossible to compute for generic mixed many-body states. In practice, the logarithmic negativity is often used to estimate the amount of bipartite entanglement~\cite{Plenio_2005, vidal_computable_2002}, which, while computable, does not have a clear operational meaning.
In fact, it is known that it fails to capture the correct `amount' of entanglement in some few-qubit examples~\cite{Wolf_2006, Tserkis_2017}. 
In this work, we highlight that many operationally meaningful measures of bipartite entanglement, such as the entanglement cost (and its exact version)~\cite{Bennett_1996, audenaert2003entanglement}, squashed entanglement~\cite{christandl2004squashed}, and distillable entanglement~\cite{Bennett_1996}, while usually intractable, can be exactly computed for a class of physically relevant strongly symmetric quantum states, which are mixtures of pure states carrying the same symmetry charge~\cite{Buca_2012, Albert_2014}. These entanglement measures can be computed for a subclass of such states, referred to as maximally mixed invariant states (MMIS),  
which naturally arise as fixed points of symmetric local open quantum evolutions~\cite{li_24, subhayan_24}.

We find that for non-Abelian symmetry algebras, which contain at least extensively (in the volume of the system) many elements, MMIS exhibit extensive and long-range entanglement coexisting with extensive classical entropy, thus enabling genuine quantum entanglement at `high temperatures'~\cite{li_23, li_24, subhayan_24}. 
Furthermore, we show that for a family of MMIS exhibiting quantum Hilbert space fragmentation (QF)~\cite{Read_2007,moudgalya_fragment_commutant_2022}, the logarithmic negativity of $\rho_{\text{QF}}$ exhibits a volume-law behavior, while the entanglement cost, distillable entanglement, and squashed entanglement all exhibit sub-volume law scaling.

Although this parametric separation provides a natural many-body example that the logarithmic negativity is not a genuine entanglement measure~\footnote{ \label{ft:footnote}A many-body \textit{pure} state with a parametric separation in entanglement entropy and logarithmic negativity was constructed in~\cite{Sherman_2016}, which was however not otherwise physically motivated.}, we find that for the SS states $\rho_{\text{QF}}$, it is equal to a version of the exact entanglement cost~\cite{audenaert2003entanglement}. Therefore, we conclude that SS states $\rho_{\text{QF}}$ can require very different entanglement resources depending on whether one seeks an exact or asymptotically-exact preparation~\footnote{A similar observation was made for a different class of states in~\cite{mori2024does}.}, revealing a specific pattern of quantum Hilbert space fragmentation.  
To explain this point, we construct a truncated state that is locally indistinguishable from $\rho_{\text{QF}}$, but which does not showcase the same parametric separation.\\


\textbf{Bipartite entanglement measures---}A density matrix $\rho_{AB}$ is entangled if it can not be represented as a mixture of product of local density matrices, $\rho_{AB} \neq \sum_{i}p_i \rho^i_A \otimes \rho^i_B$~\cite{Werner_1989}, and is separable otherwise. We can quantify the bipartite entanglement in a state in terms of the rates of transformation of the state to and from some entangled resource state (for example, EPR pairs) using standard `free' (i.e. no cost) operations in a resource theoretic sense~\cite{horodecki_quantum_2009}. Two natural sets of operations are: local operations and classical communication (LOCC) and positive partial transpose (PPT) preserving operations~\cite{Chitambar_2014}. LOCC involves all local quantum operations on $A$ and $B$, as well as unlimited classical communication between $A$ and $B$. PPT operations are a set of operations that outputs a PPT state (= a state with positive partial transpose $\rho_{AB}^{T_A}$~\footnote{The partial transpose $\rho_{AB}^{T_A}$ with respect to subsystem $A$, is given by $\langle \psi_A, \psi_B| \rho |\psi_A^\prime, \psi_B^\prime \rangle = \langle \psi_A^\prime,\psi_B|\rho^{T_B}|\psi_A, \psi^\prime_B\rangle$ for an arbitrary orthonormal basis $\{|\psi\rangle\}$ such that $|\psi\rangle = |\psi_A\rangle\otimes |\psi_B\rangle$.}) whenever acting on a PPT state. A separable mixed state is PPT, i.e. $\rho_{AB}^{T_A} \geq 0$~\cite{Peres_1996}, while the converse is not true, i.e. there are entangled states that may also be PPT~\cite{Horodecki_1998MixedState}. PPT operations include, and is strictly larger than, the set of LOCC operations, PPT $\supset$ LOCC~\footnote{For a concrete example of a PPT operation that is not LOCC, see~\cite{Bennett_1999}).}.


For a pure state, 
the bipartite entanglement is captured by the entanglement entropy, i.e. the von Neumann entropy of the reduced density matrix,
\begin{align}
    E_{(A:B)}^{\text{pure}}(\rho_{AB}) =  S(\rho_{A}) = S(\rho_{B}).
\end{align}

Operationally, entanglement can be quantified by the number of EPR pairs necessary to prepare the state using free operations, dubbed entanglement cost, or the number of EPR pairs that can be prepared from the state using free operations, which is referred to as distillable entanglement (both quantities will be precisely defined later). For pure states, both quantities are the same and equal to the entanglement entropy~\cite{Bennett_1996Concentrating}. However, this is not necessarily the case for mixed states, and the `entanglement' entropy for mixed states includes classical entropy.

\textit{Entanglement cost $E^C$} is the minimum ratio $m/n$ for the conversion of $(\rho_{\text{EPR}})^{\otimes m} \equiv \ket{\text{EPR}}\bra{\text{EPR}}^{\otimes m}$ to $\rho^{\otimes n}$ under free (LOCC or PPT) operations, with vanishing error in trace norm (denoted by $||\cdots||_1$), in the asymptotic limit $n \to \infty$ of the number of copies ~\footnote{ \label{ft:appA1}see precise definition in App.~\ref{subapp:def}}. The LOCC (or PPT) entanglement cost $E^C$ is defined as the minimum ratio $m/n$ over all choices of operations $\Phi \in {\text{LOCC (or PPT)}}$. Hereafter a measure without reference to LOCC or PPT indicates an LOCC measure, while a PPT is indicated as a superscript.

\textit{Entanglement of formation} $E^F$~\cite{Bennett_1996} is an upper bound of $E^C$:
the minimization of the averaged bipartite entanglement entropy over all (non-unique) pure state decompositions of $\rho = \sum_{i}p_i \ket{\psi_i}\bra{\psi_i}$, 
\begin{align}
    E^{F}_{(A:B)}(\rho_{AB}) = \text{min}_{\{p_i, \psi_i\}}\sum_{i}p_i S_{(A:B)}(\ket{\psi_i}\bra{\psi_i}). 
\end{align}
In fact, $E^C$ is the regularized version of the $E^F$~\cite{ hayden_asymptotic_2001}
\begin{align}
    E^{C}_{(A:B)}(\rho_{AB}) = \lim_{m \to \infty}\frac{1}{m}E^{F}_{(A:B)}(\rho_{AB}^{\otimes m}).
\end{align}

\textit{Distillable entanglement $E^D$} is the rate at which EPR pairs can be prepared from $\rho_{AB}$ via free operations~\cite{Bennett_1996Concentrating}. Consider the asymptotic setting with quantum operations applied jointly on $n$ copies $\rho_{AB}^{\otimes n}$ to distill $m$ copies of EPR pairs approximately, with the approximation error vanishing in the $n\to \infty$ limit. The maximum possible rate $m/n$ for all allowed operations is the 
distillable entanglement. Entanglement measures
$E_{(A:B)}$ that satisfy several desirable properties (monotonicity, additivity, convexity, continuity) are bounded as, $E^D_{(A:B)}\leq E_{(A:B)} \leq E^{C}_{(A:B)}$~\cite{Horodecki_2000}. One measure that satisfies all the above properties (but is nevertheless computationally intractable for a generic state) is the \textit{squashed entanglement $ E^{sq}$}, whose definition is provided in Appendix~\ref{subapp:def}. 


\indent \textit{Logarthmic negativity $E^{N}$.} Despite the powerful operational meanings of the entanglement quantities described so far, none of them is efficiently computable, as they necessarily require an optimization over mixed state decompositions, LOCC / PPT operations, or extensions of mixed states, which are computationally prohibitive (at least exponential in the size of the Hilbert space~\cite{Huang_2014}) for a generic quantum many-body system. An alternative computable entanglement measure is the \textit{logarithmic negativity $E^N$}~\cite{vidal_computable_2002, Plenio_2005},  
\begin{align}
    E^{N}_{(A:B)}(\rho_{AB}) = \log ||\rho^{T_{A}}||_{1},
\end{align}
which is non-zero for entangled states, but is not a faithful measure (i.e. it can be zero for certain entangled states as well). While the logarithmic negativity is an entanglement monotone under LOCC~\cite{Plenio_2005}, it isn't a genuine entanglement measure, as more negativity does not necessarily imply more operational entanglement \cite{Note1}. 

However, it provides an upper bound to the distillable entanglement, $E^{D} \leq E^N$~\cite{vidal_computable_2002} and is a lower bound to a stricter notion of entanglement cost, called the exact (PPT) entanglement cost~\cite{audenaert2003entanglement, Note5}. In this scenario, instead of preparing the state approximately from EPR pairs with vanishing error rates in the limit of infinite copies, we consider the rate of exactly preparing $n$ copies of the state from $m$ EPR pairs, where the error of preparation is exactly zero for a certain $m, n$ pair, before taking the asymptotic limit. It is clear that $E^{C} \leq E^{C, \text{exact}}$ for either LOCC or PPT. 
$E^N$ is upper bounded by the exact PPT entanglement cost $E^N \leq E^{C, \text{PPT, exact}}$, and the bound becomes an equality
for states which satisfy the `non-binegativity' property: $|\rho^{T_A}|^{T_A} \geq 0$, where, $|\rho| \equiv \sqrt{\rho^\dagger \rho}$~\cite{audenaert2003entanglement}.

Summarizing, these are some of the known relations between the different entanglement measures: $E^D \leq E^{sq}\leq E^{C}\leq E^{F}$, and $E^D \leq E^N \leq E^{C, \text{PPT, exact}}\leq E^{C, \text{LOCC, exact}}$ for any mixed state $\rho_{AB}$.\\

\textbf{Maximally mixed invariant state---} 
Consider a system of $L$ $N$-dits with a tensor product Hilbert space $\mathcal{H}^{(L)}$, evolving under some \emph{strongly symmetric} open quantum dynamics, e.g., via a composition of local quantum channels $\mathcal{E}_j(\bullet)=\sum_\alpha K_{\alpha, j} \bullet K_{\alpha, j}^\dagger$, or analogously, governed by a Lindblad equation. The set $\{K_{\alpha, j},K_{\alpha, j}^\dagger\}$ generates an algebra $\mathcal{A}(L)$, whose centralizer $\mathcal{C}(L)$, to which we will refer as \emph{the commutant algebra}, and that fully characterizes the strong symmetries~\footnote{A review of this formalism applied to open quantum systems can be found in Ref.~\cite{li_23}}. 
$\mathcal{A}(L)$ and $\mathcal{C}(L)$ being centralizers of each other, the Hilbert space admits the decomposition~\cite{2000_Zanardi_quantuminfo, 2007_DFS_Bartlett, 1998_lecture_von_Neumann_algebra, Read_2007}
\begin{align}\label{eq:H_decomposition}
    \mathcal{H} = \bigoplus_{\lambda \in \Lambda_L} \left[\mathcal{H}_{\lambda}^{\mathcal{C}(L)}\otimes \mathcal{H}_{\lambda}^{\mathcal{A}(L)}\right],
\end{align}
where the $\lambda$'s label the irreducible representations (irreps) of the commutant $\mathcal{C}(L)$ and the bond algebra $\mathcal{A}(L)$. Here, $\Lambda_L$ is the set of irreps compatible with a system size $L$. The irreps of the bond algebra $\mathcal{H}_{\lambda}^{\mathcal{A}(L)}$ correspond to dynamically decoupled sectors, namely Krylov subspaces, which have dimension $\text{dim}(\mathcal{H}_{\lambda}^{\mathcal{A}(L)}) = D^{(L)}_{\lambda}$. The degeneracy of these subspaces within a $\lambda$ sector is given by the dimension of the commutant irrep $\mathcal{H}_{\lambda}^{\mathcal{C}(L)}$, denoted by $\text{dim}(\mathcal{H}_{\lambda}^{\mathcal{C}(L)})  = d_\lambda$, which is $L$-independent.

Reference~\cite{li_24} recently found that when $\mathcal{C}(L)$ admits a Hopf algebra structure in the $L\to \infty$ limit, then: (i) there is a well-defined notion of a unique trivial irrep $\lambda=0$ (i.e., $d_0=1$); and (ii) given a bipartition of the system into $L=L_A+L_B $ (and for a compatible system size $L$), the subspace $\mathcal{H}_{\lambda=0}^{\mathcal{A}(L)}$ of dimension $D_0^{(L)}=\sum_{\lambda \in \Lambda_{L_A, L_B}} D_\lambda^{(L_A)}D_{\bar{\lambda}}^{(L_B)}$ admits the decomposition
$\mathcal{H}_{\lambda=0}^{\mathcal{A}(L)}=\bigoplus_{\lambda \in \Lambda_{L_A, L_B}}\left[\mathcal{H}_{\lambda}^{\mathcal{A}(L_A)}\otimes \mathcal{H}_{\bar{\lambda}}^{\mathcal{A}(L_B)}\right]$, spanned by an orthonormal basis with elements
\begin{equation} \label{eq:singlet_def}
    |\lambda; a,b\rangle = \frac{1}{\sqrt{d_\lambda}} \sum_{m=1}^{d_\lambda} \eta_{\lambda, m}|\lambda, m; a\rangle \otimes |\bar{\lambda},\bar{m};b\rangle,
\end{equation}
for $a=1,\dots, D_\lambda^{(L_A)}$, $b=1,\dots, D_{\bar{\lambda}}^{(L_B)}$, and $\eta_{\lambda, m}$ a phase. Here $\bar{\lambda}$ labels the dual representation of $\lambda$. Hence, within this invariant subspace, one finds a basis with a flat Schmidt spectrum (for the given spatial bipartition), which simplifies the calculation of various entanglement measures. 
Examples of $\mathcal{C}(L)$ with a Hopf algebra structure (in the $L\rightarrow \infty$ limit) include group algebras of finite groups, universal enveloping algebras of Lie algebras, as well as quantum groups~\cite{kassel2012quantum}.

In particular, the previous works Refs.~\cite{subhayan_24, li_24} focused on the characterization of the maximally mixed invariant state
\begin{equation}
\label{eq:MMIS_def}
    \rho_{\text{MMIS}} =\frac{1}{D_0^{(L)}}\sum_{\lambda \in \Lambda_{L_A, L_B}} \sum_{a=1}^{D_\lambda^{(L_A)}}\sum_{b=1}^{D_\lambda^{(L_B)}}|\lambda; a,b\rangle \langle \lambda; a,b|,
\end{equation}
namely the projection into the invariant subspace $\mathcal{H}_{\lambda=0}^{\mathcal{A}(L)}$. This, for example, appears as the unique stationary state of strongly symmetric open dynamics with Hermitian Kraus operators $K_{\alpha,j}=K_{\alpha,j}^\dagger$ when preparing an initial state $\rho_0$ within $\mathcal{H}_{\lambda=0}^{\mathcal{A}(L)}$. 
An illuminating example is given by an SU$(2)$ strongly symmetric evolution with local Kraus operators generated by Heisenberg terms $\textbf{S}_i\cdot \textbf{S}_{i+1}$.  
Starting from a short-range entangled state, a product of nearest-neighbor singlets, the system evolves into a highly entangled mixed stationary state as characterized by various bipartite entanglement measures~\cite{li_23, li_24, subhayan_24}.
\\


\textbf{Entanglement in the singlet ensemble---} Consider a mixture of pure states from the singlet ensemble, $\{\ket{\lambda, a, b}\}$, defined in Eq.~\eqref{eq:singlet_def}, i.e. a general density matrix of the form,
\begin{equation}
\begin{aligned}\label{eq:singlet_mixed_def}
    \rho &= \sum_{\lambda \in \Lambda_{L_A, L_B}} \sum_{a=1}^{D_\lambda^{(L_A)}}\sum_{b=1}^{D_\lambda^{(L_B)}} q_{\lambda ab}|\lambda; a,b\rangle \langle \lambda; a,b|~\\
    &\text{with}~ q_{\lambda ab} \geq 0, ~ \sum_{\lambda, a, b} q_{\lambda ab} =1, \text{and}~ p_{\lambda} \equiv \sum_{a,b} q_{\lambda ab}.\\
\end{aligned}
\end{equation}
This generalizes the $\rho_{\text{MMIS}}$ defined in Eq.~\eqref{eq:MMIS_def}, where $q_{\lambda a b} = 1/D_0^{(L)}$ for all $\lambda, a, b$. We show that all previously introduced entanglement measures can be exactly computed for these states, due to the following properties.

\begin{proposition}\label{prop:singlet}
The singlet ensemble has the following properties:
\begin{enumerate}
    \item The singlet states form a complete orthonormal basis set in the singlet subspace.
    \item{The bipartite entanglement entropy of any singlet state is given by,
    \begin{align}
        E_{A:B}(\ket{\lambda, a,b}) = \log (d_\lambda).
    \end{align}
    }
    \item {The states are locally distinguishable by LOCC (and hence by PPT operations), without destroying their bipartite entanglement.}

    \item {Any mixed state in the singlet ensemble is non-binegative, i.e. $|\rho^{T_A}|^{T_A} \geq 0$. }    
    
\end{enumerate}
\end{proposition}

These imply the following results for the bipartite entanglement structure of mixtures of singlet states: 

\begin{theorem}\label{thm:entanglement_singlets}
For a disjoint bipartition of $\rho$ in Eq.~\eqref{eq:singlet_mixed_def} on $L$ qudits, the bipartite entanglement measures are: 
\begin{enumerate}
    \item {Entanglement of formation, entanglement cost, squashed entanglement, and distillable entanglement are equal, and given by,
    \begin{align}\label{eq:Eformation}
        \mathscr{E}^{<}_{(A:B)}(\rho) = \sum_{\lambda\in \Lambda_{L_A,L_B}} p_{\lambda} \log (d_{\lambda}), 
    \end{align}
    This also implies that any mixed state entanglement measure (including squashed entanglement) $E_{A:B}$ which satisfies $E^D_{A:B}\leq E_{A:B} \leq E^{F}_{A:B}$, takes the same form.
    } 
    \item {Exact PPT entanglement cost and logarithmic negativity are given by the following expression:
    \begin{align}\label{eq:Enegativity}
        &\mathscr{E}^{>}_{(A:B)}(\rho) = \log(\sum_{\lambda\in \Lambda_{L_A,L_B}}p_\lambda d_{\lambda}).
    \end{align} }
\end{enumerate}   
\end{theorem}

The proofs of the proposition and the theorem are elementary and provided in appendices~\ref{appsec:proofs1} and ~\ref{appsec:proofs2}. \\

\textbf{Quantum fragmentation: the Temperley Lieb models---}The family of Temperley-Lieb models is the perfect ground to further investigate the distinction between various entanglement measures. This was hinted by the recent work~\cite{li_24} which analytically found a parametric separation between the entanglement negativity, the operator space entanglement, and the Rényi negativities. Here, we briefly review these models and provide an understanding of this separation in view of the analysis of the previous section and the structure of Hilbert space fragmentation. 

\begin{figure}
    \centering
    \includegraphics[width=1\linewidth]{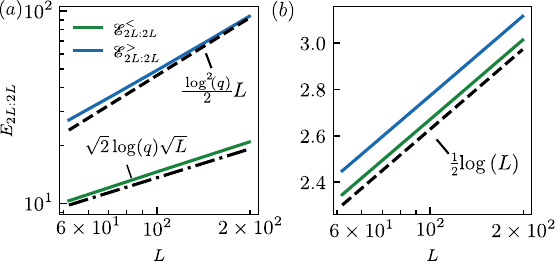}
    \caption{\textbf{Half-chain entanglement of system size $4L$ for different symmetries}. (a) Spin-$1$ Temperley-Lieb model with commutant algebra $\mathcal{C}_{\mathrm{RS}(3)}$, with $\mathscr{E}^{<}_{2L:2L} \sim \mathcal{O}(\sqrt{L})$ and $\mathscr{E}^{>}_{2L:2L} \sim \mathcal{O}(L)$, exhibiting a parametric separation. (b) Spin-$1/2$ chain with SU($2$) symmetry, where both $\mathscr{E}^{<}_{2L:2L}$ and $\mathscr{E}^{>}_{2L:2L}$ scales asymptotically as $\frac{1}{2}\log L$.}
    \label{fig:ent_comparison}
\end{figure}

Let us consider a local Hilbert space of dimension $N$ such that $\mathcal{H}^{(L)}\cong (\mathbb{C}^N)^{\otimes L}$. For these models, the bond algebra $\mathcal{A}(L)$ corresponds to the Temperley-Lieb algebra TL$_L(N)$, generated by the local operators
\begin{equation}
    e_{j,j+1}=\sum_{\alpha,\beta=1}^{N}(\ket{\alpha \alpha}\bra{\beta\beta})_{j,j+1} = |\dimer\rangle\langle\dimer|,
\end{equation}
with $|\dimer\rangle = \frac{1}{\sqrt{N}}\sum_{\alpha=1}^N|\alpha \alpha \rangle$ as a maximally entangled dimer state, and $j=1,\dots, L-1$~\cite{Read_2007, moudgalya_fragment_commutant_2022, Aufgebauer_2010}. 
These local operators satisfy the relation $e_{j,j+1}|\dimer \, \, \dimer\rangle_{i, j, j+1, k} = |\twodimer\rangle_{i,j,j+1,k}$, and the graphic notation helps to identify all Krylov subspaces~\cite{Read_2007, moudgalya_fragment_commutant_2022}.
In particular, TL$_L(2)$ is equivalently generated by Heisenberg terms $\textbf{S}_j\cdot \textbf{S}_{j+1}$, while the case $N=3$, corresponds to the SU$(3)$ symmetric couplings $(\textbf{S}_j\cdot \textbf{S}_{j+1})^2$.
The dimension of the bond algebra irreps $\mathcal{H}_\lambda^{\mathcal{A}(L)}$ is the same as the SU$(2)$ symmetric systems, given by the irreps of permutation group $S_L$ ~\cite{moudgalya_fragment_commutant_2022}.
Their commutants for all $N\geq 3$ (and a specific restriction of the system size) are the Read-Saleur commutants $\mathcal{C}_{\textrm{RS}(N\geq 3)}(L)$, which were characterized in Ref.~\cite{Read_2007}. The dimension of the irreps $\mathcal{H}_\lambda^{\mathcal{C}(L)}$ is $d_\lambda = [2\lambda+1]_{q}$, with $\lambda = 0, 1, \ldots \frac{L}{2}$ for system size $L$,  and $[n]_q = (q^{n}-q^{-n})/(q-q^{-1})$ is the $q$-deformed integer with $q\geq 1$ defined by $N=q+q^{-1}$ for $N\geq 2$.
%
%
For $N=2$, this expression recovers $d_\lambda = 2\lambda+1$ as in the SU($2$) case.
However, for $N \geq 3$, the dimension of large irreps $\lambda$ grow exponentially with $\lambda$ as $d_\lambda {\xrightarrow{\lambda \gg 1}} q^{2\lambda}$. 
Consequently, the dimension of the commutant scales exponentially with system size for $N\geq 3$. 

In recent terminology, this system is said to exhibit \emph{quantum Hilbert space fragmentation}~\cite{Moudgalya_22}. In particular, (1) the number of Krylov sectors $\sum_\lambda d_\lambda $ scales exponentially with system size; and (2) following the definition introduced in Ref.~\cite{li_23}, there is no common eigenbasis of product states for all elements in a maximal Abelian subalgebra of the commutant. Note that, as highlighted in that work, a non-Abelian commutant is not a necessary condition for quantum fragmentation, although it appears to be a sufficient condition. Nonetheless, our definition ensures that the basis on which fragmentation appears requires entanglement. \\

\textbf{Parametric separation in entanglement costs---} We now consider the maximally invariant state $\rho_{\text{MMIS}}$ defined in Eq.~\eqref{eq:MMIS_def} for the $TL_L(N)$ model, hereafter referred to as $\rho_{\text{QF}}$, on a system of length $4L$, and perform an equal spatial bipartition $L_A = L_B = 2L$. Reference~\cite{li_24} found that for large system sizes, the logarithmic negativity of $\rho_{\text{QF}}$ scales as $E_{(A:B)}^N \sim L$. However, the scaling of the $E^F$, $E^D$ and other entanglement measures (as computed in Ref.~\cite{subhayan_24} for SU$(N)$ symmetric systems), was left as an open question, which can now be easily computed using the general formulas derived in Eqs.~\eqref{eq:Eformation} and~\eqref{eq:Enegativity}.

We find that the entanglement measures such as the entanglement cost follow the scaling
\begin{equation}
\begin{aligned}
    \mathscr{E}^{<}_{2L:2L}(\rho_{\text{QF}}) \sim 2\lambda_{\text{max}}\log (q) =\log(q)\sqrt{2L} 
\end{aligned}
\end{equation}
in the limit $L\to \infty$, due to the fact that the maximum of the probability $p_\lambda$ appears at $\lambda_{\text{max}} = \sqrt{\sfrac{L}{2}}$, and $d_\lambda \sim q^{2\lambda}$ for large $\lambda$. 
On the other hand, the logarithmic negativity and the exact PPT entanglement cost result in
\begin{align}
    \mathscr{E}^{>}_{2L:2L}(\rho_{\text{QF}}) \approx \log\left(\sum_{\lambda = 0}^{2L} p_\lambda  q^{2\lambda}\right) \approx \frac{\log^2 (q)}{2}L, 
\end{align}
since the expression within the log is maximized at $\lambda_{*}  =  \frac{\log{(q)}}{2}L$. The distinct asymptotic analysis for the different entanglement measures is visualized in Fig.~\ref{fig:TL_cutoff}, and derived in detail in Appendix~\ref{appsec:asymptote}.

Thus the logarithmic negativity and the exact PPT entanglement cost scale as volume law $\sim L$, while the entanglement of formation and distillation, squashed entanglement, and all other good entanglement measures scale as $\sim \sqrt{L}$. In particular, this implies that there is a parametric separation in the resources that are necessary to prepare $\rho_{\text{QF}}$ asymptotically rather than exactly (via PPT operations). We can understand this point more clearly by considering a truncated state that is arbitrarily close to $\rho_{\text{QF}}$, but whose exact preparation requires parametrically different number of entangled resources.\\

\begin{figure}
    \centering
    \includegraphics[width=0.95\linewidth]{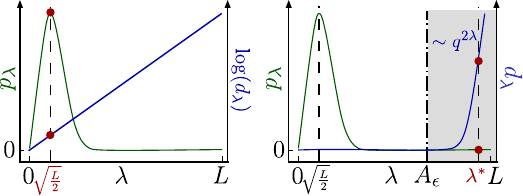}
    \caption{\textbf{The weight of irreps $p_\lambda$ and the dimension $d_\lambda$ of the irreps of $\mathcal{C}_{\mathrm{RS}(N\geq3)}(4L)$ at half-chain bipartition.} 
    (a) The weight of irreps $p_\lambda$ peaks at $\lambda_{\mathrm{max}}\approx \sqrt{\frac{L}{2}}$, and is vanishingly small with large $\lambda$, which gives the scaling of $\mathscr{E}_{2L:2L}^{<} \sim \log q^{2\lambda_{\mathrm{max}}} \sim \sqrt{L}$.
    (b) The dimension $d_\lambda$ scales exponentially with $\lambda$ as $q^{2\lambda}$, leading to the scaling $\mathscr{E}^{>}_{2L:2L} \sim L$, as dominated by the term $\lambda = \lambda_*$ (see main text).
    The truncated state $\rho^{\epsilon}_{\text{QF}}$ is  constructed by truncating $p_\lambda$ for $\lambda\geq A_{\epsilon}$. While $\rho^{\epsilon}_{\text{QF}}$ is locally distinguishable from $\rho_{\text{QF}}$, both its $\mathscr{E}^{>}_{2L:2L}$ and $\mathscr{E}^{<}_{2L:2L}$ scale as $\mathcal{O}(\sqrt{L})$} 
    \label{fig:TL_cutoff}
\end{figure}


\textbf{Locally indistinguishable truncated states---} Consider a state $\rho_{\text{QF}}^{\epsilon}$ constructed by truncating large irreps $\lambda > A_\epsilon$, such that $\sum_{\lambda=A_{\epsilon}}^{\text{min}(L_A, L_B)} p_{\lambda} = \epsilon\ll 1$, for $\epsilon$ not scaling with system size. 
For an equal bipartition of the system and for any $\epsilon$, we can always choose $A_{\epsilon}$ to be a multiple of $\lambda_{\text{max}}$, i.e. $A_{\epsilon} = a_{\epsilon} \sqrt{\sfrac{L}{2}}$~\footnote{This can be seen by explicitly computing the integral using the asymptotic form of the $p_{\lambda}$ given in Eq.~\ref{eq:p_asymp}, $\int_{a_{\epsilon}\sqrt{L/2}}^{\infty} p_{\lambda} d\lambda \approx (2a_{\epsilon}e^{-a_{\epsilon}^2}+\sqrt{\pi} \text{Erfc}(a_{\epsilon}))/\sqrt{\pi}$, which can be made as small as necessary by choosing large enough $a_\epsilon$.}. This leads to a suppression in the trace distance between the truncated and the original state, ${||} \rho_{\text{QF}}^{\epsilon} - \rho_{\text{QF}}{||}_{1} \approx 2\epsilon$,
which implies that for any bounded operator $\hat{O}$, the difference of expectation values is only $\epsilon$ away $|\textrm{tr}(\rho_{\text{QF}}\hat{O})-\textrm{tr}(\rho^{\epsilon}_{\text{QF}}\hat{O})|\leq 2\epsilon ||\hat{O}||_\infty$.

The truncated state is also a mixture of states from the singlet ensemble; hence we use Eq.~\eqref{eq:Enegativity} to compute the exact PPT entanglement cost in the asymptotic limit $L \to \infty$,
which becomes $\mathscr{E}^{>}_{2L:2L}(\rho^{\epsilon}_{\text{QF}}) \approx a_{\epsilon}\log(q)\sqrt{2L}$.
However, $\mathscr{E}^{<}_{2L:2L}(\rho^{\epsilon}_{\text{QF}})$ computed as per Eq.\eqref{eq:Eformation}, showcases the same scaling as before truncating the state , i.e., $\mathscr{E}^{<}_{2L:2L}(\rho^{\epsilon}_{\text{QF}})\sim \log(q)\sqrt{2L}$.
Therefore, we find a parametric separation between the exact PPT entanglement costs of the exact and truncated states $\rho_{\text{QF}}$ and $\rho^{\epsilon}_{\text{QF}}$, which as we previously found, scales linarly with system size $L$. Additional details can be found in App.~\ref{app:local_indis}.\\

\textbf{Conclusions---} In this work, we exactly computed various bipartite entanglement measures for a class of long-range entangled many-body mixed states arising from strongly symmetric open quantum dynamics, with special focus on quantum Hilbert space fragmentation. 
Symmetric mixed states often lead to simplified expressions of entanglement~\cite{Vollbrecht_2001}; however the fact that these entanglement measures can be computed for the strongly symmetric states for arbitrary non-commuting symmetry algebras is by itself a novel result, and deserves further attention. We end with a few open questions motivated by our results.

Our work highlights the pattern of quantum fragmentation for Temperley-Lieb algebras, demonstrating a parametric separation among various entanglement measures. Is this result general for systems showcasing quantum fragmentation~\cite{23_Brighi_particle_conserving_east, 24_Chen_quantum_fragmentation_breakdown, 2021_minimal_HSF_with_local_constraints, 2024_Parameswaran_SY_commutant}? Here we have focused on bipartite entanglement, but it is also interesting to probe their multipartite entanglement structure.

Another natural question is whether such exotic behaviors of entanglement persist for small (both symmetric and non-symmetric) perturbations of the maximally mixed invariant state, which can be regarded either as steady states of certain strongly symmetric open systems, or as infinite temperature states satisfying the symmetry constraints. Characterizing the dynamics of the various entanglement measures in such symmetric open dynamics (which is computationally hard), or the entanglement structure of strongly symmetric Gibbs states of local Hamiltonians at finite temperature, are interesting future directions to pursue.

\section*{Acknowledgments}
We thank Olexei Motrunich for inspiring discussions regarding finite temperature Gibbs states, and Leonardo Lessa for comments on the draft. S.S. thanks Amin Moharramipour, Leonardo Lessa, Chong Wang, and Timothy Hsieh for collaboration in related previous work. Y.L. and P.S. thank Frank Pollmann and Nicholas Read for collaboration in related previous work. P.S. acknowledges support from the Caltech Institute for Quantum Information and Matter, an NSF Physics Frontiers Center (NSF Grant No.PHY-1733907), and the Walter Burke Institute for Theoretical Physics at Caltech. Y.L. acknowledges the DFG Germany’s Excellence Strategy–EXC–2111–390814868.  Research at PI is supported in part by the Government of Canada through the Department of Innovation, Science and Economic Development Canada and by the Province of Ontario through the Ministry of Colleges and Universities. 



%

\appendix
\section{Proofs and computations}\label{sec:proofs}
In this section we complete the proofs and computations in the paper and also provide some exact expressions and definitions of the various entanglement measures.

\subsection{Definitions of entanglement measures} \label{subapp:def}
The entanglement cost and the exact entanglement cost defined in the main text are given by the following expressions,
\begin{align}
    E^C_{(A:B)}(\rho_{AB}) = &\inf_{r}\{r: \lim_{n\to \infty}  \nonumber \\
    &\left(\inf_{\Phi}  ||\rho_{AB}^{\otimes n} - \Phi(\rho_{\text{EPR}}^{\otimes rn})||_{1}\right) = 0 \}\\
    E^{C, \text{exact}}_{(A:B)}(\rho_{AB}) = &\lim_{n\to \infty}\inf_{r_n}\{r_n: \nonumber \\ & \left(\inf_{\Phi}   ||\rho_{AB}^{\otimes n} - \Phi(\rho_{\text{EPR}}^{\otimes r_n n})||_{1}\right) = 0 \},
\end{align}
where the optimization is done over all allowed operations $\Phi \in \text{LOCC or PPT}$, depending on the context.

\textit{Squashed entanglement $ E^{sq}$} is the minimized conditional mutual information for any allowed extensions of $\rho_{AB}$ and satisfies most of the desired properties for an entanglement measure~\cite{ christandl2004squashed, Brand_o_2011},
\begin{align}
    E^{sq}_{(A:B)}(\rho_{AB}) &= \inf_{\rho_{ABC}}\left\{ \frac{1}{2}I(A:B|C): 
 \Tr _C \rho_{ABC} = \rho_{AB}\right\},  \nonumber \\
 \text{ where, } &I(A:B|C) = S_{AC} +S_{BC} - S_{ABC} - S_C.
\end{align}

To define logarithmic negativity, we need to compute the partial transpose on $A$, which is defined with respect to a specific basis $\{\ket{a}\}$ in $A$ and general basis $\{\ket{b}\}$ in $B$ as follows,
\begin{align}
    \rho_{AB} &= \sum_{a,a', b,b'} \left[\rho_{AB}\right]_{a,a'; b,b'} \ket{a}\bra{a'}\otimes \ket{b}\bra{b'}, \nonumber\\
    \rho^{T_A}_{AB} &\equiv \sum_{a,a', b,b'} \left[\rho_{AB}\right]_{a,a'; b,b'} \ket{a}\bra{a'}\otimes \ket{b'}\bra{b}.
\end{align}

\subsection{Proofs of Proposition~\ref{prop:singlet}}\label{appsec:proofs1}
Statements 1 and 2 follow directly from the decomposition 
\begin{equation}
    \mathcal{H}_{\lambda=0}^{\mathcal{A}(L)}=\bigoplus_{\lambda \in \Lambda_{L_A, L_B}}\left[\mathcal{H}_{\lambda}^{\mathcal{A}(L_A)}\otimes \mathcal{H}_{\bar{\lambda}}^{\mathcal{A}(L_B)}\right],
\end{equation}
spanned by the orthonormal basis in Eq.~\eqref{eq:singlet_def}, and also shown in~\cite{li_24, subhayan_24}. Statement 3 was proven in~\cite{subhayan_24} and adapted from~\cite{livine_entanglement_2005}. It is implied by the decomposition of the singlet states in Eq.~\eqref{eq:singlet_def} as follows: Alice and Bob with access to $A$ and $B$ can measure the labels $\lambda, a$ and $\overline{\lambda}, b$ respectively by local projective measurements, and share the measurement information by classical communication. This measurement does not destroy the superposition over $m$ labels in Eq.~\eqref{eq:singlet_def} - thus the entanglement between $A$ and $B$ is preserved. We can also prove the non-binegativity of the singlet ensemble states (statement 4) using the convenient bipartite decomposition in Eq.~\eqref{eq:singlet_def}, which leads to,
\begin{align}
   \rho &= \sum_{\lambda,a,b} q_{\lambda ab}\sum_{m,n} \ket{\lambda, m; a}\bra{\lambda, n; a} \otimes \ket{\overline{\lambda}, m; b}\bra{\overline{\lambda}, n; b} \nonumber\\ 
   \rho^{T_A} &= \sum_{\lambda,a,b} q_{\lambda ab}\sum_{m,n} \ket{\lambda, n; a}\bra{\lambda, m; a} \otimes \ket{\overline{\lambda}, m; b}\bra{\overline{\lambda}, n; b} 
\end{align}
Now, to compute $|\rho^{T_A}|$, we need to use the fact that the orthonormal basis states have the property, $\bra{\lambda, n; a}\lambda',n'; a'\rangle = \delta_{\lambda\lambda'}\delta_{nn'}\delta_{aa'}$ on both partitions. Using this fact, and the fact that each of the $\lambda, a, b$ sectors are orthogonal, you get,
   \begin{align}
       |\rho^{T_A}| &= \sum_{\lambda,a,b} q_{\lambda ab}\sum_{m,n} \ket{\lambda, n; a}\bra{\lambda, n; a} \otimes \ket{\overline{\lambda}, m ;b}\bra{\overline{\lambda}, m; b}. 
   \end{align}
   Now notice, that $|\rho^{T_A}|$ is a valid quantum state which is also separable, and hence PPT. Thus $\rho$ is not bi-negative.

\subsection{Proofs of Theorem~\ref{thm:entanglement_singlets}}\label{appsec:proofs2}
Statement 1 was proved in~\cite{subhayan_24}, and we provide a sketch of the proof here. Consider the mixed state defined in Eq~\eqref{eq:singlet_mixed_def} - the averaged entanglement of this particular decomposition in terms of singlet states is an upper bound to the entanglement of formation $E^F$, 
\begin{align}
    E^F \leq \sum_{\lambda, a, b} q_{\lambda ab}\log d_\lambda.
\end{align}
Furthermore, Alice and Bob can measure $\lambda, a, b$ labels and distil the pure state $\ket{\lambda,a,b}$ by LOCC (see the proof of the Proposition~\ref{prop:singlet}). This LOCC strategy leads to a \textit{lower bound} to the distillable entanglement which is the same as the RHS of the equation above. Using the fact that $E^D \leq E^F$, and the other relations between the entanglement measures listed in the main text, we must have that $E^F = E^C = E^{sq} = E^D = \mathscr{E}^{<}_{(A:B)}(\rho) = \sum_{\lambda, a, b} q_{\lambda ab}\log d_\lambda$.

Statement 2 can be proved by explicitly computing the logarithmic negativity for a bipartition of the state in~\eqref{eq:singlet_mixed_def}, which was done in~\cite{subhayan_24, li_24}. Next, we use the property of non-binegativity of the singlet ensemble and the results from~\cite{audenaert2003entanglement} to arrive at the result that the exact PPT entanglement cost is given by 
\begin{align}
    &\mathscr{E}^{>}_{(A:B)}(\rho) = \log\left(\sum_{\lambda, a, b}q_{\lambda ab} d_{\lambda}\right) = \log\left(\sum_{\lambda\in \Lambda_{AB}}p_\lambda d_{\lambda}\right).
\end{align} 
\subsection{Asymptotic analysis of bipartite entanglement in $TL_L(N)$}\label{appsec:asymptote}

The dimension of the bond algebra irreps $\mathcal{H}_\lambda^{\mathcal{A}(L)}$ for the Temperley Lieb models is the same as the dimension of the Krylov subspaces of the SU$(2)$ symmetric systems, and are given by 
\begin{equation}
\begin{aligned}
     D_\lambda^{(L)}
    =\frac{2\lambda +1}{L/2+\lambda +1}\begin{pmatrix}L\\
    L/2+\lambda
    \end{pmatrix}.
\end{aligned}
\end{equation}

To analytically obtain these scalings, we approximate
$p_\lambda\equiv \frac{D^{(2L)}_{\lambda} D^{(2L)}_{\overline{\lambda}} }{D^{(4L)}_{0}}$
by the following asymptotic expression using Stirling's approximation 
\begin{align}\label{eq:p_asymp}
 p_\lambda \approx \frac{8\sqrt{2}}{\sqrt{\pi}}\frac{\lambda^2}{L^{3/2}}e^{-2\lambda^2/L},
\end{align}
for $1\ll\lambda \ll L$, which makes the calculation of both the entanglement of formation and logarithmic negativity of $\rho_{\text{MMIS}}$ straightforward. First notice that $\lambda$ takes values in $\{0, 1, \cdots, L\}$. The probability is maximized at $\lambda_{\text{max}} = \sqrt{\frac{L}{2}}$. Thus, the entanglement of formation as well as the entanglement cost, and therefore all good mixed entanglement measures follow the scaling 
\begin{equation}
\begin{aligned}
    \mathscr{E}^{<}_{2L:2L}(\rho_{\text{QF}}) \sim 2\lambda_{\text{max}}\log (q)=\log(q)\sqrt{2L}.  
\end{aligned}
\end{equation}
in the limit $L\to \infty$, where we have used the fact that $d_\lambda \sim q^{2\lambda}$ for large $\lambda$. On the other hand, the logarithmic negativity and the exact PPT entanglement cost $E_{(A:B)}^{C, \text{PPT, exact}}$) are given by a slightly different expression, which requires a different asymptotic analysis,
\begin{align}
    \mathscr{E}^{>}_{2L:2L}(\rho_{\text{QF}}) \approx \log\left(\sum_{\lambda = 0}^{2L} p_\lambda  q^{2\lambda}\right) 
\end{align}

The expression within the log is dominated by the maximum of the exponential factor $e^{-2\lambda^2/L}q^{2\lambda} = e^{-2\lambda^2/L + 2\lambda \log q}$, which is maximized at $\lambda_{*}  = L \log{q}/2$. This leads to the asymptotic expression,
\begin{align}
    \mathscr{E}^{>}_{2L:2L}(\rho_{\text{QF}}) &= \frac{\log^2 q}{2}L  +O(\sqrt{L}).
\end{align}

\section{Proof of Local indistinguishibility} \label{app:local_indis}
Let us define the state $\rho_{\text{QF}}^{\epsilon}$ by truncating large irreps from $\rho_{\text{QF}}$ (for a given bipartition), defined with respect to the probability distribution (see Fig.~\ref{fig:TL_cutoff}),
\begin{align}\label{eq:plam_eps}
    p^{\epsilon}_{\lambda} = \begin{cases}
   &\frac{p_{\lambda}}{1-\epsilon} \text{ for } 0 \leq \lambda \leq A_{\epsilon} \\ 
   & 0 \text{ otherwise,}
   \end{cases}
\end{align}
where $A_\epsilon$ is defined such that $\sum_{\lambda=A_{\epsilon}}^{\text{min}(L_A, L_B)} p_{\lambda} = \epsilon$, where $\epsilon$ is chosen to be a small non-negative number, $\epsilon \ll 1$. 
For the equal bipartition of the system and for any $\epsilon$, we can always choose $A_{\epsilon}$ to be a multiple of $\lambda_{\text{max}}$, i.e. $A_{\epsilon} = a_{\epsilon} \sqrt{\frac{L}{2}}$. This can be seen by explicitly computing the sum using the asymptotic form of the $p_{\lambda}$ given in Eq.~\eqref{eq:p_asymp}, $\sum_{\lambda >A_{\epsilon}\sqrt{L/2}} p_{\lambda}\approx \int_{a_{\epsilon}d\lambda \, \sqrt{L/2}}^{\infty} p_{\lambda} \approx (2a_{\epsilon}e^{-a_{\epsilon}^2}+\sqrt{\pi} \text{Erfc}(a_{\epsilon}))/\sqrt{\pi}$, which can be made as small as necessary by choosing large enough $a_\epsilon$. This leads to a suppression in the trace distance between the truncated and the original state,
\begin{align}
    &{||} \rho_{\text{QF}}^{\epsilon} - \rho_{\text{QF}}{||}_{1} = \sum_{\lambda} |p_{\lambda} - p_{\lambda}^{\epsilon}| \approx 2\epsilon, 
\end{align}
which implies that for any bounded operator $\hat{O}$, the difference of expectation values $|\textrm{tr}(\rho_{\text{QF}}\hat{O})-\textrm{tr}(\rho^{\epsilon}_{\text{QF}}\hat{O})|\leq ||\rho_{\text{QF}}-\rho^{\epsilon}_{\text{QF}}||_1||\hat{O}||_\infty$ is only $\epsilon$ away.

The truncated state is also a mixture of states from the singlet ensemble; hence we use Eq.~\eqref{eq:Enegativity} to compute the exact PPT entanglement cost in the asymptotic limit $L \to \infty$,
\begin{align}
    \mathscr{E}^{>}_{2L:2L}(\rho^{\epsilon}_{\text{QF}}) = \log{\left(\sum_{0}^{A_{\epsilon}}p^{\epsilon}_{\lambda}d_{\lambda}\right)} \approx a_{\epsilon}\log(q)\sqrt{2L}.
\end{align}

However, $\mathscr{E}^{<}_{2L:2L}(\rho^{\epsilon}_{\text{QF}})$ computed as per Eq.\eqref{eq:Eformation}, showcases the same scaling as before truncating the state , i.e., $\mathscr{E}^{<}_{2L:2L}(\rho^{\epsilon}_{\text{QF}})\sim \log(q)\sqrt{2L}$. Therefore, we find the following parametric separation between the exact PPT entanglement costs of the exact and truncated states $\rho^{\epsilon}_{\text{QF}}$ and $\rho_{\text{QF}}$,
\begin{align}
    &E^{C, \text{PPT, exact}}(\rho_{\text{QF}})-E^{C, \text{PPT, exact}}(\rho^{\epsilon}_{\text{QF}}) \approx \frac{\log^2 (q)}{2}L.
\end{align}

\end{document}